*Classification*: Physical Sciences: Engineering

*Title*: **Characterization of reconnecting vortices in superfluid helium**

*Authors*:
Gregory P. Bewley[1], Matthew S. Paoletti[1], Katepalli R. Sreenivasan[1,2], and Daniel P. Lathrop[1],

[1]University of Maryland, College Park, MD 20740
[2]International Center for Theoretical Physics, Trieste, Italy, 34014

*Corresponding author*:
Katepalli R. Sreenivasan
The Abdus Salam International Centre for Theoretical Physics
Strada Costiera 11
34014 Trieste, Italy
Email: krs@ictp.it
Phone: (+39) 040 2240 251
Fax: (+39) 040 2240 410

*Manuscript information*:
9 pages, 4 figures, 0 tables
128 word abstract (250 max)
26,905 character paper (47,000 max)

*Abstract*:

**When two vortices cross, each of them breaks into two parts and exchanges part of itself for part of the other. This process, called vortex reconnection, occurs in classical as well as superfluids, and in magnetized plasmas and superconductors. We present the first experimental observations of reconnection between quantized vortices in superfluid helium. We do so by imaging micron-sized solid hydrogen particles trapped on quantized vortex cores (Bewley GP, Lathrop DP, Sreenivasan KR, 2006, *Nature*, 441:588), and by inferring the occurrence of reconnection from the motions of groups of recoiling particles. We show the distance separating particles on the just-reconnected vortex lines grows as a power law in time. The average value of the scaling exponent is approximately ½, consistent with the scale-invariant evolution of the vortices.**

**Introduction**

Vorticity in superfluid helium is confined to filaments only angstroms in diameter, about which the fluid circulates with quantized angular momentum (1). The reconnection of two such quantized vortices can occur when their cores come into contact, as illustrated in Fig. 1. When reconnection occurs, each core breaks at one point into two halves and exchanges half of itself for half of the other. After reconnection the vortices draw away from each other. Although this process is thought to be an essential feature of superfluid turbulence (2, 3), and in other systems mentioned below, it has never been observed in helium until now. This paper is a first report of such observations.

While we cannot directly observe quantized vortex lines, we infer their locations by observing the motions of micron-sized solid hydrogen particles. We have previously demonstrated that these particles can be trapped on quantized vortex cores (4, 5). Although not all the particles are trapped by vortices, sequences of images reveal that some particles are arranged along curves, and that these curves possess properties expected of quantized vortices. When two of these curves cross, their subsequent evolution is consistent with the reconnection dynamics. In particular, we study the distance between recoiling vortices, and find that this distance evolves as expected of vortices just after reconnection. This is our tool for studying the reconnection of quantized vortices.

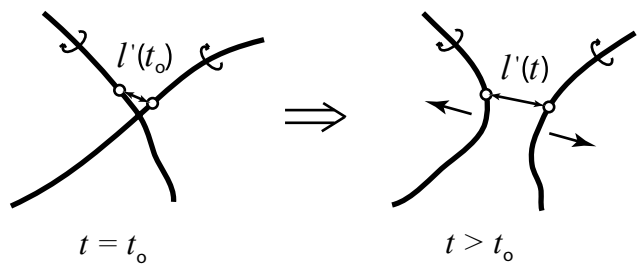

**Fig. 1.** The illustration shows the cores of reconnecting vortices at the moment of reconnection at $t_0$, and after reconnection. The small circles mark the particles trapped on the vortex cores. The arrows indicate the motion of the trapped particles and vortices recoiling due to their large curvature. We measure the distance between two particles over time.

Quantized vortices can be considered theoretically as phase singularities and as topological defects in the order parameter describing the superfluid. In that sense, analogues to the vortices exist in a wide range of systems where reconnection is also an essential feature, and where our work has possible implications. These systems include superconductors (6), liquid crystals (7), and heart tissue (8). In addition, reconnection is thought to play an important role in the dynamics of magnetic field lines in magnetized plasmas (9), where it affects solar convection and space weather. While reconnection is difficult to observe in many of these systems, it has been observed directly in a Newtonian fluid (10,11) and in liquid crystals (7). In superfluid helium, the phenomenon has been studied using numerical simulations of the nonlinear Schrödinger equations (12,13), and this study provided evidence that the picture of superfluid turbulence as consisting of reconnecting vortices (2,3) is essentially correct. The evolution of vortices after reconnection has also been explored using line-vortex models (14,15), and described analytically (16,17). Below, we compare our results to the predictions of these studies, and to the following argument.

As a simple model of quantized vortex dynamics, assume that the evolution of two interacting vortices is scale-invariant in time, $t$, and can be characterized by a single length scale, $l(t)$. Such a characterization is possible when the influences of the initial and boundary conditions are negligible (18), and if $l(t)$ is much larger than the vortex diameter. It then follows from dimensional arguments that a single parameter, $\kappa t / l^2(t)$, describes the evolution. Here, we assume that the only relevant material parameter is the quantum of circulation of the superfluid vortex, $\kappa = h/m$, where $h$ is Planck's constant, and $m$ is the mass of a helium atom. According to Buckingham's $\pi$ theorem (19), if a physically meaningful equation exists relating $l(t)$ and $t$, then $l(t)$ must be proportional to $(\kappa t)^{1/2}$. Numerical simulations using the Biot-Savart law to describe the evolution of line vortices also find a $t^{1/2}$ dependence of the distance between reconnecting vortices (15), as does the analysis of Ferrell (16) for the radius of curvature of the vortices. The above argument can be applied to characterize the motions of other phenomena such as the evolution of a vortex ring.

In the context of just-reconnected vortices, we propose that the minimum distance between the cores of vortices recoiling after reconnection is an adequate measure of the scale parameter, $l(t)$. The characterization of the dynamics as scale-invariant is then possible when the following conditions hold: the initial radius of curvature of the vortices is much larger than the largest $l(t)$ under consideration, the largest $l(t)$ is much smaller than the distance to other vortices in the system, the timescale of large-scale distortions of the fluid is much larger than the largest time under consideration, and the smallest $l(t)$ under consideration is much larger than the vortex core diameter. We revisit these conditions in light of the data below.

**Experiment**

The experiments are performed in a cryostat with a rectangular $5 \times 5 \times 25$ $cm^3$ channel containing liquid helium-4, which was continually replenished. The long axis of the channel was vertical, and a window in each of the vertical faces had an optical aperture of 2.5 $cm$. A laser beam, formed into a sheet passes through one pair of windows, illuminates particles in the helium. The laser power varies between 1 and 5 $W$, the sheet thickness between 100 and 500 µm, and its width between 4 and 8 $mm$. A CMOS movie camera gathers the light scattered at 90 degrees from the illuminating volume with a resolution of 16 µm per pixel. We collect data at either 20 or 40 frames per second, and with 50 $ms$ or 25 $ms$ exposure times, respectively. The camera captures the positions of particles within the illuminating volume projected onto a plane.

We form hydrogen particles according to the method described in Bewley et al. (4,5), which is to inject a room-temperature mixture of hydrogen and helium gases into liquid helium at a temperature just above the lambda-point, which is the superfluid phase-transition temperature. The gas mixture freezes into nearly neutral particles, whose diameter is estimated to be on the order of a *μm*. The volume fraction of the particles is about $10^{-5}$. When the fluid is cooled, using a mechanical vacuum pump, and temperatures below the lambda-point are reached, superfluid vortices collect particles along their cores and that the particles appear as evenly spaced dots. The observed lines, corresponding to temperatures from 2.01 *K* to 2.14 *K*, are about 130 *μm* apart on the average (20).

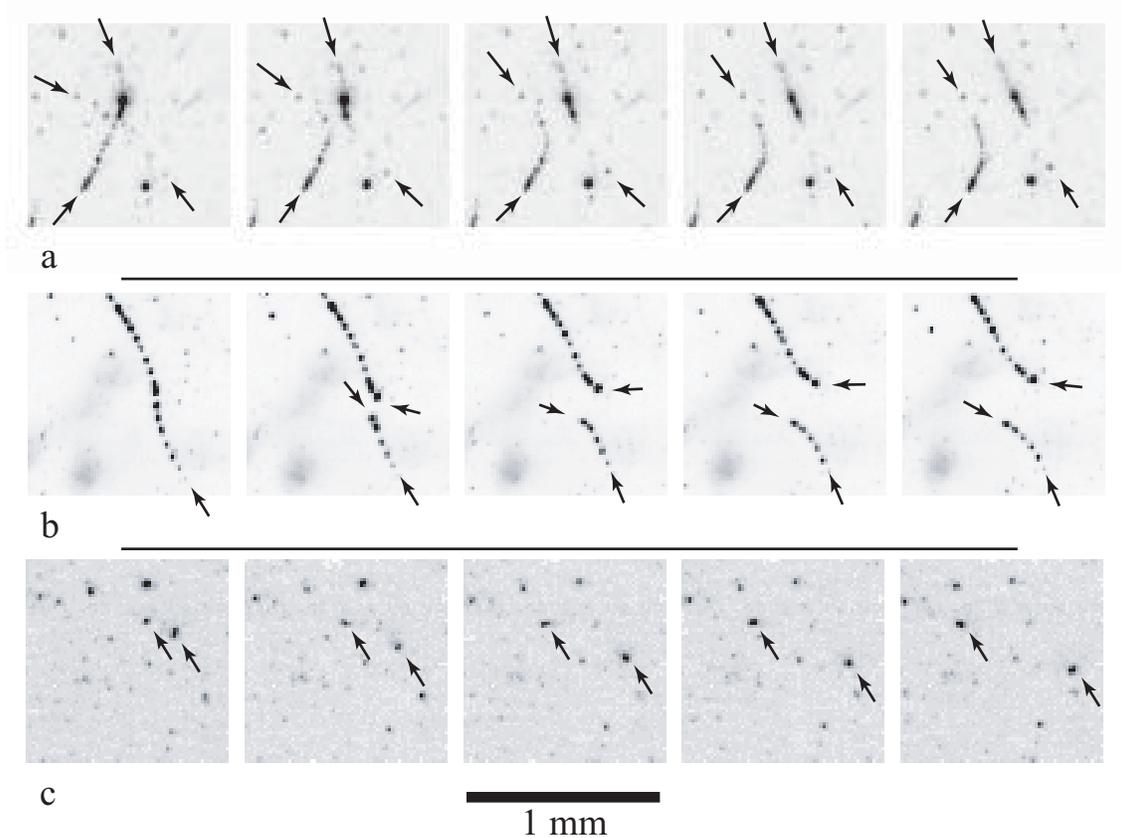

1 mm

**Fig. 2.** Images of reconnecting vortices in superfluid helium, made visible by particles trapped on the vortex cores. Each series of frames in (a), (b) and (c) are images of hydrogen particles suspended in liquid helium, taken at 50 *ms* intervals. Some of the particles are trapped on quantized vortex cores, while others are randomly distributed in the fluid. Arrows show where the decorated vortex cores appear to leave the illuminating light sheet and become invisible. Before reconnection, the particles drift collectively with the background flow in a configuration similar to that shown in the first frames of (a), (b) and (c). Subsequent frames show reconnection as the sudden motion of a group of particles. In (a), both vortices participating in the reconnection have several particles along their cores. In projection, the approaching vortices in the first frame appear crossed. In (b), particles initially make only one vortex visible, the other vortex probably has not yet trapped any particles. In (c), we infer the existence of a pair of reconnecting vortices from the sudden motion of a single pair of particles recoiling from each other.

We steadily vary the cooling rate between 0.1 *mK/s* and 0.4 *mK/s*, with a mean value of 0.22 *mK/s*. Over the time that we follow a single pair of reconnecting vortices, the change in the energy of the vortices due to the change in temperature is small so that the system is in a quasi-steady state: we estimate the fractional change in the density of the superfluid, $\Delta\rho_s/\rho_s$, over the period of observation of each reconnection (about 1 *s*) to be about 0.4% at 2.14 *K*, using the properties given by Donnelly (1).

We characterize the resultant flow using particle image velocimetry (PIV) (21). Despite the complicated interactions between individual particles and quantized vortices (22), the use of PIV is justified since the majority of particles move collectively, and along smooth trajectories. The root-mean-square velocity of the flow within the field of view, *u*, varied from one realization to another between 0.1 *mm/s* and 0.4 *mm/s*, with a mean value of 0.2 *mm/s*. We measure the characteristic length scale, *L*, by computing the integral of the longitudinal correlation function averaged over the data from all realizations, and found it to be 7 *mm*. Further methods of analysis of the data, including particle tracking, are described below.

**Results**

We record instances in which particles behave in a way that is qualitatively consistent with their being trapped on reconnecting vortices (Fig. 2). That is, we seek instances where at least two particles move apart suddenly against a background of relatively motionless particles. We interpret the motions of such particles in the framework of superfluid vortices undergoing reconnection as in Fig. 1.

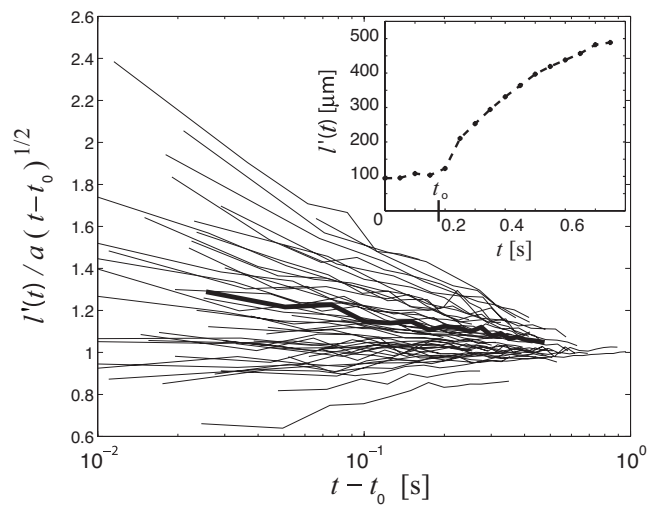

**Fig. 3.** The inset shows the measurements, $l(t)$, for a typical candidate reconnection. Before reconnection occurred the pair of particles traveled together, and we estimated the time of reconnection, $t_0$, as described below. The curves in the main figure are the measurements divided by $a(t-t_0)^\beta$ for 52 candidate reconnections, where *a* and $t_0$ were found by fitting the measurements to the power law with arbitrary scaling exponent, $\beta$. If the data were governed by perfect power laws, each curve would follow a line that passes through the point (1, 1). If the scaling exponents were all ½, the data would lie on a horizontal line. A randomly chosen curve is plotted with a thicker line in order to show the path of a typical single trajectory.

We select for consideration particle images according to the following criteria. First, while reviewing the images of flow we noticed abrupt motions, or large accelerations, of pairs of particles signaling an apparent reconnection. The velocities of these particles were larger after reconnection than before. In most cases, at least one of the vortices had a few particles along its core, and there was no evidence that other vortices were nearby. In other words, we expect that vortices not participating in the reconnection were at least more distant than the maximum separation between reconnected vortices, although we cannot say this with certainty because of the limited thickness of the illuminating sheet. Second, if the initial radius of curvature of one of the vortices was small, we ignored the event. For each event satisfying the above criteria, we located, by hand, the pair of particles that were abruptly moving away from each other and nearest to the reconnection point. The locations of the particles were estimated with sub-pixel accuracy by fitting the particle images to a Gaussian function and finding the peak of the Gaussian function. We determine the particle positions for each image frame for as long as the particle pair was visible after reconnection.

Data analysis proceeds as follows. Reconnection of two vortices occurs at some time, $t_0$, but is invisible until the recoil of the vortices propagates far enough to affect the particles. The initial separation of these particles is a measure of their distance from the reconnection point, and was 85 $\mu m$ on the average. We make two assumptions. First, once the particles have responded to the reconnection, the distance between the pair of particles closest to the reconnection point, $l'(t)$, is thereafter a measure of the scale parameter, $l(t)$. Second, the scale parameter evolves as a power law, $l(t) = a(t-t_0)^\beta$. We estimate the time origin of the reconnection, $t_0$, as well as the scaling exponent, $\beta$, and the amplitude of the event, $a$, in the following way. We perform a linear least-squares fit of the logarithm of the data to a first-order polynomial, to obtain values of $a$ and $\beta$, for a series of values of $t_0$. We then select the $t_0$ that minimizes the mean square of the differences between the data and the power law.

We recorded 52 candidate reconnections and on average captured each with 13 observations of particle position pairs, spanning about 1.5 decades in time. In Fig. 3, we plot the data, $l(t)$, compensated for a power law with a scaling exponent of ½, and using the estimates $t_0$ and $a$. The figure shows that individual trajectories are nearly power laws, whose scaling exponents often deviate from ½. As shown in Fig. 4, we find a distribution of scaling exponents whose mean value is 0.45, and whose standard deviation is 0.07. The mean value of the amplitude of reconnections, $a$, is about 718 $\mu m/s^\beta$; if $\beta$ were equal to ½, then $a$ would be about 2.3 times larger than $\kappa^{1/2}$.

Also shown in Fig. 4 is the distribution of scaling exponents for the distance separating pairs of initially proximal particles that are chosen arbitrarily from the background, using the same sequences of images as those chosen for reconnection studies. 353 pairs of tracks that were nearby at some moment during their trajectory were studied. The mean scaling exponent for these random particle pairs is zero, which reflects the fact that the background flow evolves over longer time scales and on larger length scales than do the reconnections. We are aware that the moderately negative values of the exponent in the background flow suggest a loss of differentiability, but we are prevented from commenting on this property because, the choice of the virtual origin imposes a large uncertainty when exponent is close to zero. We interpret the main conclusion of the figure to imply that the background flow has an exponent close to zero while those involving reconnections have an exponent centered near 0.45. We cannot confirm that its difference from ½ is significant.

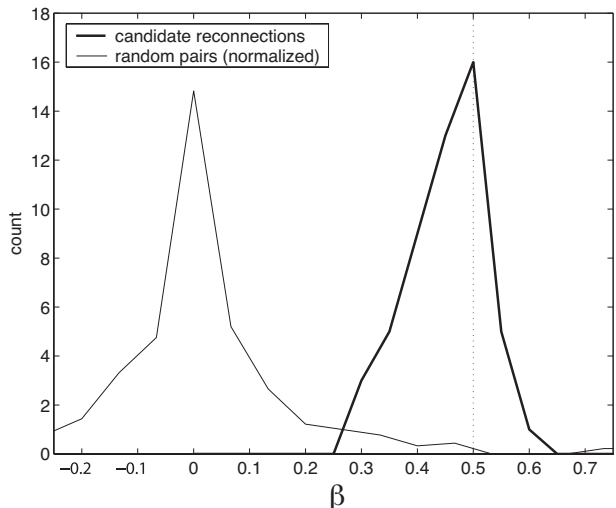

**Fig. 4.** Histogram of the scaling exponents for the data in Fig. 3, as well as those found for randomly chosen particle trajectories, as described in the text. The histogram for the random-pair data has the same area beneath it as the histogram for the experimental data. The mean value of the scaling exponents for the candidate reconnections is about 0.45, indicated by the vertical dotted line.

**Discussion**

Several phenomena may affect the evolution of reconnecting vortices. These include strain rates present in the superfluid motion, dynamic pressure effects, finite-size effects, and the initial curvature of the vortices and angle between them. In addition, the presence of particles on the vortex cores may modify the behavior of the vortices by the action of viscous drag on the particles, particle inertia, or the stabilization of vortex intersections by the particles. Finally, mutual friction (1) may add damping to the vortex motion, which is not included in the simple model used to derive the ½-power. We discuss some of these effects below, although we do not know if they are sufficient to cause the small deviation observed from the expected scaling.

The conditions described at the beginning of the paper concerning the asymptotic state hold only approximately in the experiment, and this approximation may introduce a bias into our measurement of the scaling exponent. Among the deviations from the asymptotic state is the large-scale distortion of the flow. We estimate the characteristic time for these distortions to be $L/u = 35\ s$, where $L$ and $u$ are given above. This time is much larger than the time over which we observe each reconnection, so we expect its effect to be small.

The following argument suggests that the effect of mutual friction, or damping, on the observed motion of quantized vortices is weak. One way to gauge the effect of damping is to examine the behavior of waves on a vortex core, known as Kelvin waves (1). Such waves on quantized vortices are underdamped at any temperature in liquid helium. One can then compare the period of a wave on a vortex to the timescale of its decay. This ratio is $q = \alpha/(1-\alpha')$, where $\alpha$ and $\alpha'$ are the two parameters describing the strength of mutual friction, and where a smaller value of $q$ indicates weaker damping (23). For the range of temperatures of our observations, all below 2.14 $K$, $q$ is less than one and varies between 0.67 and 0.29, indicating that the effect of mutual friction is small.

It is interesting to note that numerical simulations of the Biot-Savart law (15) reveal a $t^{1/2}$ dependence of the scale parameter even before reconnection, whereas we observe no such dependence before reconnection. This may be because before reconnection, the motions of vortices on the scales observed were dominated by large-scale distortions of the fluid, until very short times before reconnection, which were probably not observed.

In conclusion, we capture motions of particles trapped on quantized vortices and follow their motion through reconnection events. The time dependence of the separation distance between the particles closest to the point of reconnection is close to that predicted by dimensional analysis, as well as by the numerical simulation of line-vortices. A consequence of this scaling is that the velocities near the time and point of reconnection are large. For example, if the measured scaling holds when the vortices are about two core diameters from each other, the vortices would be moving away from each other at about 220 $m/s$. Of course, we cannot resolve distances on these scales, and it is unlikely that micron-sized hydrogen particles would faithfully trace such motions. While there is considerable use of particle tracking in classical fluids to study dynamics, the response of particles to superfluid motion is still not completely understood. We believe the observations presented in this paper will be useful for interpreting the motions of particles in superfluid turbulence, and will therefore be useful for understanding superfluid turbulence itself. In addition, new theoretical progress may be needed to understand trapped particles and their precise role near reconnection. Finally, we hope these observations may serve as useful conceptual models for reconnecting phenomena in the analogous systems involving smooth fluid vortices or magnetic field lines in plasmas or superconductors.

## Acknowledgements

We thank Tom Antonsen, Russ Donnelly, Chris Lobb, Ed Ott, Nigel Goldenfeld, and Makoto Tsubota for discussions, and are grateful for the financial support of the National Science Foundation (Division of Materials Research), NASA, and the Center for Nanophysics and Materials at the University of Maryland.